\documentclass[conference,10pt]{IEEEtran}
\def\ps@headings{%
\def\@oddhead{\mbox{}\scriptsize\rightmark \hfil \thepage}%
\def\@evenhead{\scriptsize\thepage \hfil \leftmark\mbox{}}%
\def\@oddfoot{}%
\def\@evenfoot{}}
\makeatother

\usepackage{cite}
\usepackage{url}
\usepackage{simplemargins}

\usepackage[dvips]{graphicx}
\usepackage{color}
\usepackage{url}
\usepackage{graphicx}
\usepackage{color}
\usepackage{url}

\usepackage{amssymb}
\usepackage{times}
\usepackage{theorem}
\usepackage{subfigure}

\input{epsf}

\newcommand {\beq} {\begin{equation}}
\newcommand {\eeq} {\end{equation}}
\newcommand {\barr} {\begin{array}}
\newcommand {\earr} {\end{array}}
\newcommand {\bear} {\begin{eqnarray}}
\newcommand {\eear} {\end{eqnarray}}
\newcommand {\bears} {\begin{eqnarray*}}
\newcommand {\eears} {\end{eqnarray*}}

\newcommand{\eat}[1]
{}

\newtheorem{thm}{Theorem}[section]
\newtheorem{assumption}{Assumption}[section]

\newtheorem{proposition}{Proposition}

\newtheorem{remark}{Remark}
\newtheorem{corollary}{Corollary}
\newtheorem{definition}{Definition}

\def\R{\mathop{\rm I\kern -0.20em R}\nolimits}

\def\b1{{\bf 1}}

\begin{document}

\title{From Altruism to Non-Cooperation in Routing Games }

\author{Amar Prakash Azad$^{*+}$ , Eitan Altman$^*$ and R. El-Azouzi$^+$\\
$^*$ Maestro group, INRIA, 2004 Route des Lucioles, Sophia Antipolis, France \\
$^+$LIA, University of Avignon, 339, chemin des Meinajaries,
Avignon, France\\
\{amar.azad,eitan.altman\}@sophia.inria.fr,
rachid.elazouzi@univ-avignon.fr. }

\maketitle

\begin{abstract}

The paper studies the routing in the network shared by several
users. Each user seeks to optimize either its own performance or
some combination between its own performance and that of other
users, by controlling the routing of its given flow demand. We
parameterize the degree of cooperation which allows to cover the
fully non-cooperative behavior, the fully cooperative behavior, and
even more, the fully altruistic behavior, all these as special cases
of the parameter's choice. A large part of the work consists in
exploring the impact of the degree of cooperation on the
equilibrium. Our first finding is to identify multiple Nash
equilibria with cooperative behavior that do not occur in the
non-cooperative case under the same conditions (cost, demand and
topology). We then identify Braess like paradox (in which adding
capacity or adding a link to a network results in worse performance
to all users) in presence of user's cooperation. We identify another
type of paradox in cooperation scenario: when a given user increases
its degree of cooperation while other users keep unchanged their
degree of cooperation, this may lead to an improvement in
performance of that given user. We then pursue the exploration and
carry it on to the setting of Mixed equilibrium (i.e. some users are
non atomic-they have infinitesimally small demand, and other have
finite fixed demand). We finally obtain some theoretical results
that show that for low degree of cooperation the equilibrium is
unique, confirming the results of our numerical study.

\end{abstract}

\section{Introduction}

Non-cooperative routing has long been studied both in the framework
of road-traffic as well as in the framework of telecommunication
networks. Such frameworks allow to model the flow configuration that
results in networks in which routing decisions are made in a
non-cooperative and distributed manner between the users. In the
case of a finite (not very large) number of agents, the resulting
flow configuration corresponds to the so called Nash equilibrium
\cite{BsrOlsd} defined as a situation in which no agent has an
incentive to deviate unilaterally. The Nash equilibrium has been
extensively used in telecommunications, see e.g.
\cite{orda,altman_cdc}. The authors in \cite{orda} studied a routing
games in which each user has a given amount of flow to ship and has
several paths through which he may split that flow. Such a routing
game may be handled by models similar to \cite{PSC99} in the special
case of a topology of parallel links. This type of topology is
studied in detail in the first part of \cite{orda} as well as in
\cite{AlKa00}. However, the model of \cite{PSC99} does not extend
directly to other topologies. Indeed, in more general topologies,
the delay over a \textit{path} depends on how much traffic is sent
by other users on any other path that shares common links. Routing
games with general topologies have been studied, for example, in the
second part of \cite{orda}, as well as in \cite{AlKa00}. A related
model was studied thirty years ago by Rosenthal in
\cite{Rosenthal73}, yet in a discrete setting.
It is
shown that in such a model there always exists a pure strategy Nash
equilibrium. He introduces a kind of discrete potential function for
computing the equilibrium. Nevertheless if a player has more than 1
unit to ship such an equilibrium doesn't always exist.

In this work, we embark on experimental investigation of the impact
of cooperation in the context of routing games. 
In particular we consider parallel links and load balancing network
topology for investigation, originally presented in \cite{orda} and
\cite{kameda_altman} in the context of selfish users. The
experimentation is mainly aimed at exploring some strange behaviors
which appears in presence of user's partial cooperation (Cooperation
in Degree), which is further strengthened with some theoretical
results.

Firstly, we identify loss of uniqueness of Nash equilibria. We show
by a simple example of parallel links and load balancing network
that there may exist several such equilibria. Moreover, even the
uniqueness of link utilization at equilibria may fail even in the
case of simple topology. A similar example of parallel links, in
absence of the cooperation between users there would be a single
equilibrium \cite{orda}. Beyond Nash equilibrium we investigate
further in the setting of Mixed users i.e. where there are two types
of users, Group user and Individual users. Group users seek Nash
equilibrium while the Individual users seek equilibrium with Wardrop
conditions. Strengthening our earlier finding, we observe loss of
uniqueness with partial cooperation against the unique solutions
shown in \cite{mixed_eq}  for selfish users.  However in the latter
section (Sec. \ref{sec:unique}), we show theoretically that there
exist uniqueness of Nash equilibrium under some conditions in the
presence of cooperation between users.

Secondly, we identify paradoxical behavior in presence of such
cooperation. One of the observed paradox here is a kind of Braess
paradox, a well studied paradox in routing context. Braess paradox
has attracted attention of many researchers in context of routing
games especially related to upgrading the system, see
\cite{Bean}-\cite{kameda_altman}. The famous Braess paradox tell us
that increasing resources to the system leads to degraded
performance in some cases. Such paradox is originally shown to exist
in many scenarios, e.g. Braess network in \cite{kameda_infocom2001},
Load balancing network in \cite{kameda_altman}. Although such
paradoxes are found even in the case of selfish users earlier, their
existence even in case of such partial cooperation is highlighted
here. We show that as the link capacity increases the overall cost
of a user decreases i.e. addition of resources in the system can
tentatively lead to degraded performance. Even more, we also
identify another kind of paradox related to \emph{degree of
Cooperation}: When a user increases its degree of cooperation while
other users keep their degree of cooperation unchanged, leads to
performance improvement of that user. We also observe similar
behavior even when other user also increase their degree of
cooperation.

The paper is structured as follows : In section
\ref{sec:system_model},  we present the system model, define our
framework of cooperative user and, formulate the problem. Further in
section \ref{sec:num_inv} we detail the numerical investigation and
summarize the findings. Based on one of the findings, we depict more
examples identifying Braess paradox in the setting of Nash game in
subsection  \ref{subsec:bras_exa}. In section \ref{sec:mixed_eq},
mixed equilibrium is illustrated. In section  \ref{sec:unique}, we
develop theoretical results to show the conditions where uniqueness
can be
established in presence of users cooperation. 
In section \ref{sec:conclusion} we summarize the study of impact of
cooperation.

\section{System model}
\label{sec:system_model}

We consider a network ${\cal( V,L )}$, where $\cal{V}$ is a finite
set of nodes and ${\cal L \subseteq V \times V } $ is a set of
directed links. For simplicity of notation and without loss of
generality, we assume that at most one link exists between each pair
of nodes (in each direction). For any link $ l =(u, v) \in {\cal L}$
,define $S(l) = u$ and $D(l) = v$. Considering a node $v \in {\cal
V}$, let In$(v) = \{l : D(l) = v \}$ denote the set of its in-going
links, and Out$(v) = \{l : S(l) = v\}$ the set of its
out-going links. 

A set ${\cal I} = \{ 1, 2,..., I\}$ of users share the network $\cal
(V,L)$, where each source node acts as a user in our frame work. We
shall assume that all users ship flow from source node $s$ to a
common destination $d$. Each user $i$ has a throughput demand that
is some process with average rate $r^i$. User $i$ splits its demand
$r^i$ among the paths connecting the source to the destination, so
as to optimize some individual performance objective. Let $f_l^i$
denote the expected flow that user $i$ sends on link $l$. The user
flow configuration $\mathbf{f}^i = {(f^i_l)}_{l\in{\cal L}}$ is
called a routing strategy of user $i$. The set of strategies of user
$i$ that satisfy the user's demand and preserve its flow at all
nodes is called the strategy space of user $i$ and is denoted by
${\bf F}^i$, that is:


 \bears {\bf F}^i =\{\mathbf{f}^i \in
\mathbb{R}^{|{\cal L}|} ; \sum_{l\in \textrm{Out}(v)} f^i_l =
\sum_{l\in \textrm{In}(v)} f^i_l+r^i_v,v\in {\cal V} \},\eears where
$r^i_s = r^i, r^i_d = -r^i$ and $r^i_v= 0$ for $v\neq s, d$. The
system flow configuration $\mathbf{f} = (f^1,...,f^I)$ is called a
\emph{routing strategy profile} and takes values in the product
strategy space ${\bf F} = \otimes_{i\in {\cal I}} {\bf F}^i$.

The objective of each user $i$ is to find an admissible routing
strategy ${\bf f}^i\in {\bf F}^i$ so as to minimize some performance
objective, or cost function, $J^{i}$, that depends upon ${\bf
f}^{i}$ but also upon the routing strategies of other users. 
Hence $J^i({\bf f})$ is the cost of user $i$ under routing strategy
profile ${\bf f}$.

\subsection{Nash equilibrium}
Each user in this frame work minimizes his own cost functions which
leads to the concept of Nash equilibrium. The minimization problem
here depends on the routing decision of other users, i.e., their
routing strategy \bears\label{nash_equilibrium}
\textbf{f}^{-i}=(\textbf{f}^{1},...,\textbf{f}^{i-1},\textbf{f}^{i+1},...\textbf{f}^{I}),
\eears
\begin{definition}
A vector ${\mathaccent "7E {\bf f}}^i$, $i=1,2,...,I$ is called a
Nash equilibrium if for each user $i$, ${\mathaccent "7E {\bf f}}^i$
minimizes the cost function given that other users' routing
decisions are ${\mathaccent "7E {\bf f}}^{j}$, $j\not=i $. In other
words, \bear\label{eq:Nash_def} {J}^i({\mathaccent
"7E {\bf f}}^1, {\mathaccent "7E {\bf f}}^2,..., {\mathaccent "7E
{\bf f}}^I) =\min_{{\bf f}^i\in {\bf F}^i} {J}^i({\mathaccent "7E
{\bf f}}^1, {\mathaccent "7E {\bf f}}^2,..., {\bf f}^i,...,
{\mathaccent "7E {\bf f}}^I),&&\nonumber\\ i=1,2,...,I,&& \eear
where ${\bf F}^i$ is the routing strategy space of user $i$.
\end{definition}

Nash equilibrium has been discussed in the context of
non-cooperative game with selfish users quite often in recent
studies.

In this paper we study a new aspect of cooperative routing games
where some users cooperate with the system taking into account the
performance of other users. We define this \emph{degree of
Cooperation} as follows :

{}
\begin{definition}\label{sec:coop}
Let $\overrightarrow{\alpha ^i}= (\alpha ^i_1,..,\alpha ^i_{|{\cal
I}|})$ be the \emph{degree of Cooperation} for user $i$. The new
operating cost function $\hat{J}^i$ of user $i$ with Degree of
Cooperation, is a convex combination of the cost of user from set
${\cal I }$, \bears \hat{J}^i(\mathbf{f})= \sum_{k\in {\cal I}}
\alpha^i_k J^k(\mathbf{f}) ; \,\, \sum_k \alpha^i_k =1
,i=1,...|{\cal I}|\eears where $\hat{J}^i(\textbf{f})$ is a function
of system flow configuration $\textbf{f}$ with cooperation.
\end{definition}
Based on the \emph{degree of Cooperation} vector, we can view the
following properties for user $i$,
\begin{itemize}
\item Non cooperative user : if $\alpha^i_i=0 $.
\item Altruistic user : User $i$ is fully cooperative with all users and does not care for his
benefits, i.e., $\alpha^{i}_i=1 $.
\item Equally cooperative - if $\alpha^i_j=\frac{1}{|{\cal{P}}|} $, user $i$
is equally cooperative with each user $j$, where $j \in {\cal P},
{\cal P} \subseteq {\cal I} $.
\end{itemize}
Note that the new operating cost function of a user is the
performance measure with \emph{degree of Cooperation}, where it
takes into account the cost of other users. Although a user
cooperating with the system, it attempts to minimize its own
operating cost function in the game setting. Hence such frame work
can be classified under non-cooperative games and the thus we can
benefit to apply the properties of non-cooperative games in such
scenario.

\section{Numerical Investigation of the role of cooperation}
\label{sec:num_inv} In this section we detail some numerical
examples to study the routing game in the presence of cooperation
between some users. In these examples, we use two types of cost
functions : linear function which is often used in the
transportation network and M/M/1 function which is used in the
queuing networks. We consider two network topologies : parallel
links \cite{orda} and load balancing networks \cite{altman_cdc}
which are defined below

\begin{figure}[tb]
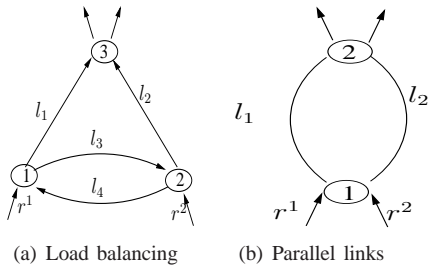
{}
\begin{center}
{}\subfigure[Load balancing]{
          \label{fig:lb}
          \resizebox{2.5cm}{3cm}{\input{2user.pstex_t}}}
{} \subfigure[Parallel links]{
         \label{fig:para_link}
\resizebox{2.5cm}{3cm}{         \input{para11.pstex_t}}}
\caption{Network Topology}
\end{center}{}
\end{figure}

\emph{Load Balancing Network}: A simple load balancing topology of
network ${\cal G}$ consists of $3$ nodes is depicted in Fig.
\ref{fig:lb}. This topology has been widely studied in context of
queuing networks. The nodes are numbered $1,2,3$ and communication
links among them are numbered as $l_1,l_2,l_3,l_4$. Node $1,2$ acts
as source node and node $3$ acts as destination node. Link $l_1,l_2$
are directed links for nodes $1,3$ and nodes $2,3$ where as, link
$l_3,l_4$ are directed link for nodes $1,2$ and nodes $2,1$. Cost
function of user $i$ is the sum of cost of each link $J^i=\sum_{l\in
\{1,...4\}} f^i_l T_l(f_l)$ , where $T_l(f_l)$ is the link cost
function. The cost of each user $i$ with cooperation can be defined
as below, \bear \hat{J}^i &=& \sum_{l \in \{ 1,...4\}} \sum_{k\in \{
1,2\}} \alpha^i_k f^k_l T_l(f_l) \eear

\emph{Parallel Links Network}: A simple parallel links topology of
network ${\cal G}$ consists of $2$ nodes is depicted in Fig.
\ref{fig:para_link} which is originally discuses in \cite{orda}. The
nodes are numbered $1,2$ and communication links between them are
numbered as $l_1,l_2$. Node $1$ acts as source node and node $2$
acts as destination node. 
Cost function of user $i$ is the
sum of cost of each link $J^i=\sum_{l\in \{1,2\}} f^i_l T_l(f_l)$ ,
where $T_l(f_l)$ is the link cost function. The cost of each user
$i$ with cooperation can be defined as below, \bear \hat{J}^i &=&
\sum_{l \in \{ 1,2\}} \sum_{k\in \{ 1,2\}} \alpha^i_k f^k_l T_l(f_l)
\eear For each network topology, we consider both the cost functions
for investigation.

 \emph{Linear Cost Function}: Linear link cost
function is defined as, $T_l(f_{l_i})= a_i f_{l_i}+ g_i $ for link
$i=1,2$, where as, $T_l(f_{l_j})= c f_{l_j}+ d $ for link $j=3,4$.

\emph{M/M/1 Delay Cost Function}: The link cost function can be
defined as, $T_l(f_{l_i})= \frac{1}{C_{l_i}-f_{l_i}} $ , where
$C_{l_i}$ and $f_{l_i}$ denote the total capacity and total flow of
the link $l_i$. Note that this cost represents the average expected
delay in a  M/M/1 queue with exponentially distributed inter arrival
times and service times  under various regimes such as the FIFO
(First In First Out) regime in which customers are served in the
order of arrivals, the PS (Processor sharing) regime and the LIFO
(Last In First Out) regime. This same cost describes in fact the
expected average delays in other settings as well such as the M/G/1
queue (exponentially distributed inter arrival times and general
independent service times) under the PS or the LIFO regime.
%
\subsection{Numerical Examples}
\label{sec:num_ex} \label{sec:num_ex} 
\begin{figure}[tb]
{}\subfigure[Cost at NEP ]{ \label{fig:nashc0:J}
\includegraphics[width=.4\textwidth]{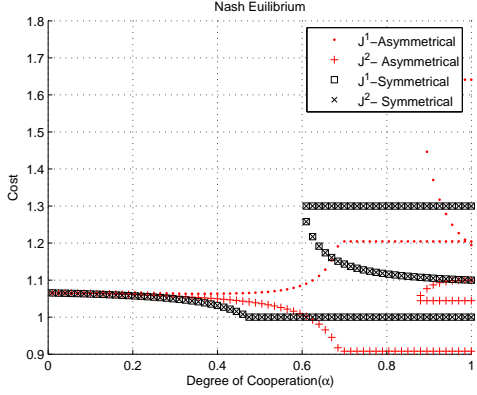}}
{}\hfill \subfigure[Flow values at NEP ]{ \label{fig:nashc0:x}
\includegraphics[width=.4\textwidth]{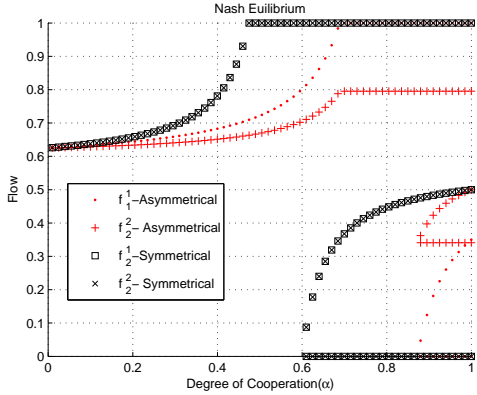}}
{}\caption{Topology : Load balancing, Cost function : Linear,
Parameters : $a=1,c=0,d=0.5$, Cooperation : \{ Symmetrical:
$\alpha^1=\alpha^2$, Asymmetrical: $0\leq\alpha^1\leq1, \alpha^2=0
$\}.}\label{fig:nash_1} \label{fig:nashc0}
\end{figure}

We consider two users share a
network. We distinguish two cases. An asymmetric case in which the
user 1 is cooperative with $\alpha^1_1>0$ and user 2 is
noncooperative, i.e., $\alpha^2_2=0$. The second case is symmetric
case in which both users are cooperative with the same degree of
cooperation $\alpha$. We compute the Nash equilibrium at
sufficiently many points of \emph{degree of Cooperation} $\alpha$ in
the interval [0,1] and plot the corresponding user cost and user
flow. Here user flow signifies the fraction of demand flowing in the
corresponding user destination link. Since we consider only two
links, the fraction of demand flow in one route complements that of
the other route. Hence we plot the fraction of demand corresponding
to the user, i.e., $f^1_{l_1}$ for user $1$ and $f^2_{l_2}$ for user
$2$. In sequel we describe five experiments as follows:

{\emph{Experiment 1) Load balancing network with linear link cost}}:
 In Fig. \ref{fig:nashc0:J}-\ref{fig:nashc0:x}, we plot the cost and
the flow obtained at Nash equilibrium versus $\alpha$ in the range
[0, 1]. Note that the plot of user $1$ and $2$ overlap in the figure
in symmetrical case. This is due to the same degree of Cooperation.

\begin{figure}[tb]
{}\subfigure[Cost function at Nash equilibrium ]{
\label{fig:nashc2:J}
\includegraphics[width=.4\textwidth]{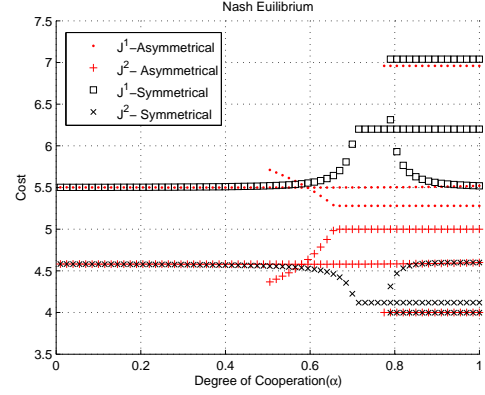}}
{}\hfill \subfigure[Flow values at Nash equilibrium ]{
\label{fig:nashc2:x}
\includegraphics[width=.4\textwidth]{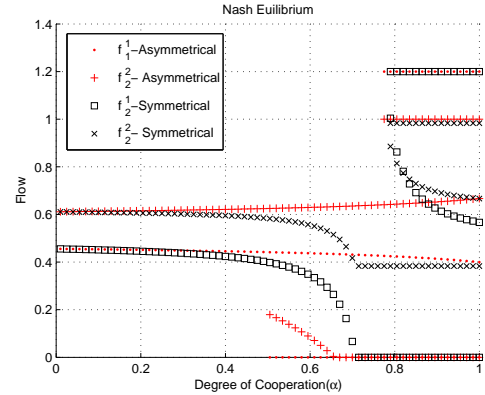}}
{}\caption{Topology : Parallel links, Cost function : Linear,
Parameters : $a=1,c=0,d=0.5$, Cooperation: \{ Symmetrical:
$\alpha^1=\alpha^2$, Asymmetrical: $0\leq\alpha^1\leq1, \alpha^2=0
$\}.}\label{fig:nashc2}
\end{figure}

{\emph{Experiment 2) Parallel links with linear link cost}}:
 In Fig. \ref{fig:nashc2:J},\ref{fig:nashc2:x}, we plot the cost
function and the flow for both users obtained at Nash equilibrium
for $\alpha$ in the range of [0, 1].

\begin{figure}[tb]
{}\subfigure[Cost function at Nash equilibrium ]{
\label{fig:nashc6:J}
\includegraphics[width=.4\textwidth]{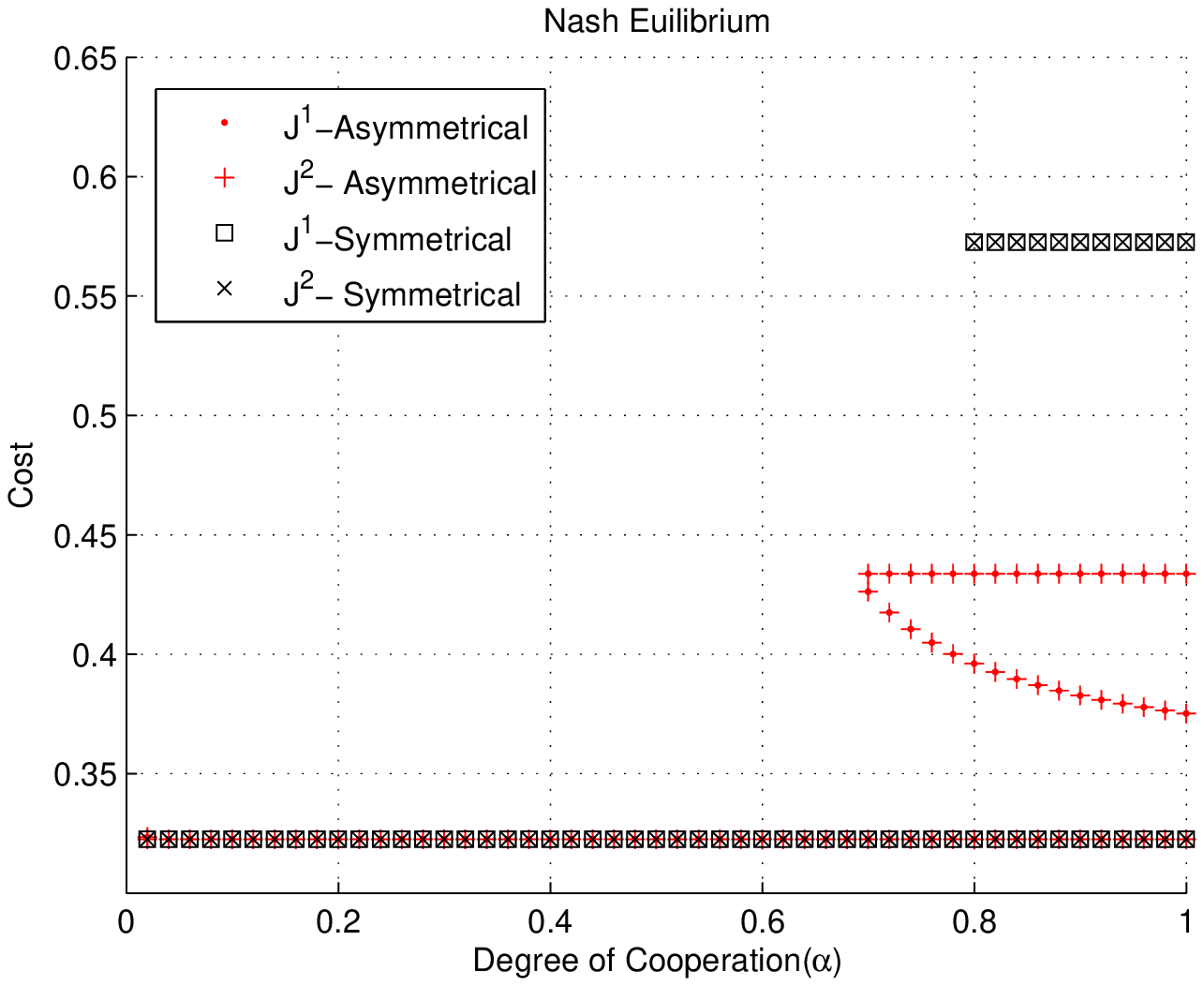}}
{}\hfill \subfigure[Flow values at Nash equilibrium ]{
\label{fig:nashc6:x}
\includegraphics[width=.4\textwidth]{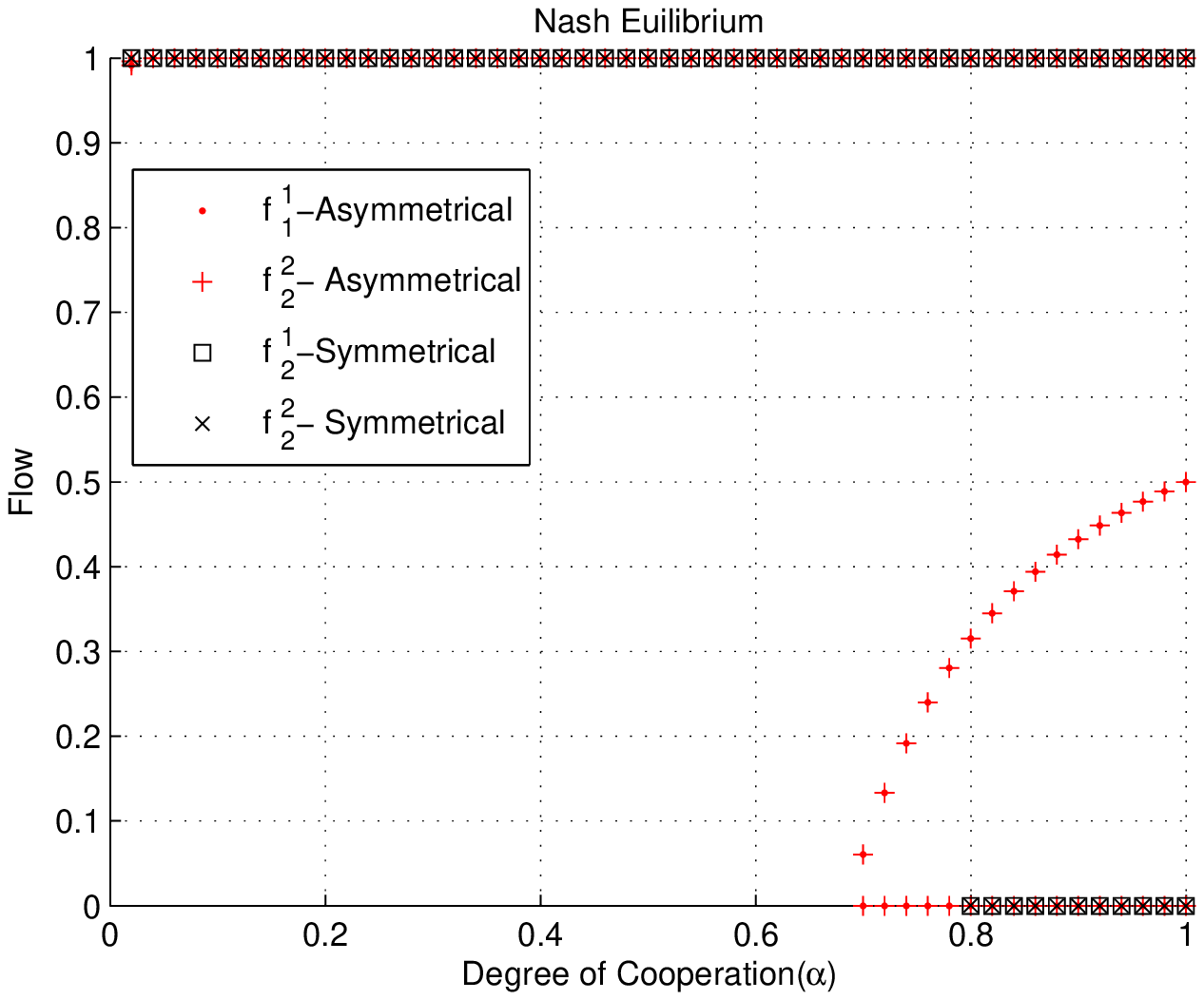}}
{}\caption{Topology : Load balancing, Cost function : M/M/1 Delay,
Parameters : $C_{l_1} = 4.1, C_{l_2} = 4.1, C_{l_3} = 5, C_{l_4} =
5, r^1 = 1, r^2 = 1$, Cooperation: \{ Symmetrical:
$\alpha^1=\alpha^2$, Asymmetrical: $0\leq\alpha^1\leq1, \alpha^2=0
$\}.}\label{fig:nashc6}
\end{figure}

{\emph{ Experiment 3) Load balancing network with M/M/1 link cost}}:
Consider the parameters for the link cost functions as, $a_1=4 ,
g_1=1 ,a_2=2 ,g_2=2, r^1=1.2, r^2=1 $. In Fig.
\ref{fig:nashc6:J},\ref{fig:nashc6:x}, we plot cost and flow
obtained at Nash equilibrium for  $0\leq \alpha\leq 1$.

\begin{figure}[tb]
{}\subfigure[Cost function at Nash equilibrium ]{
\label{fig:nashc4:J}
\includegraphics[width=.4\textwidth]{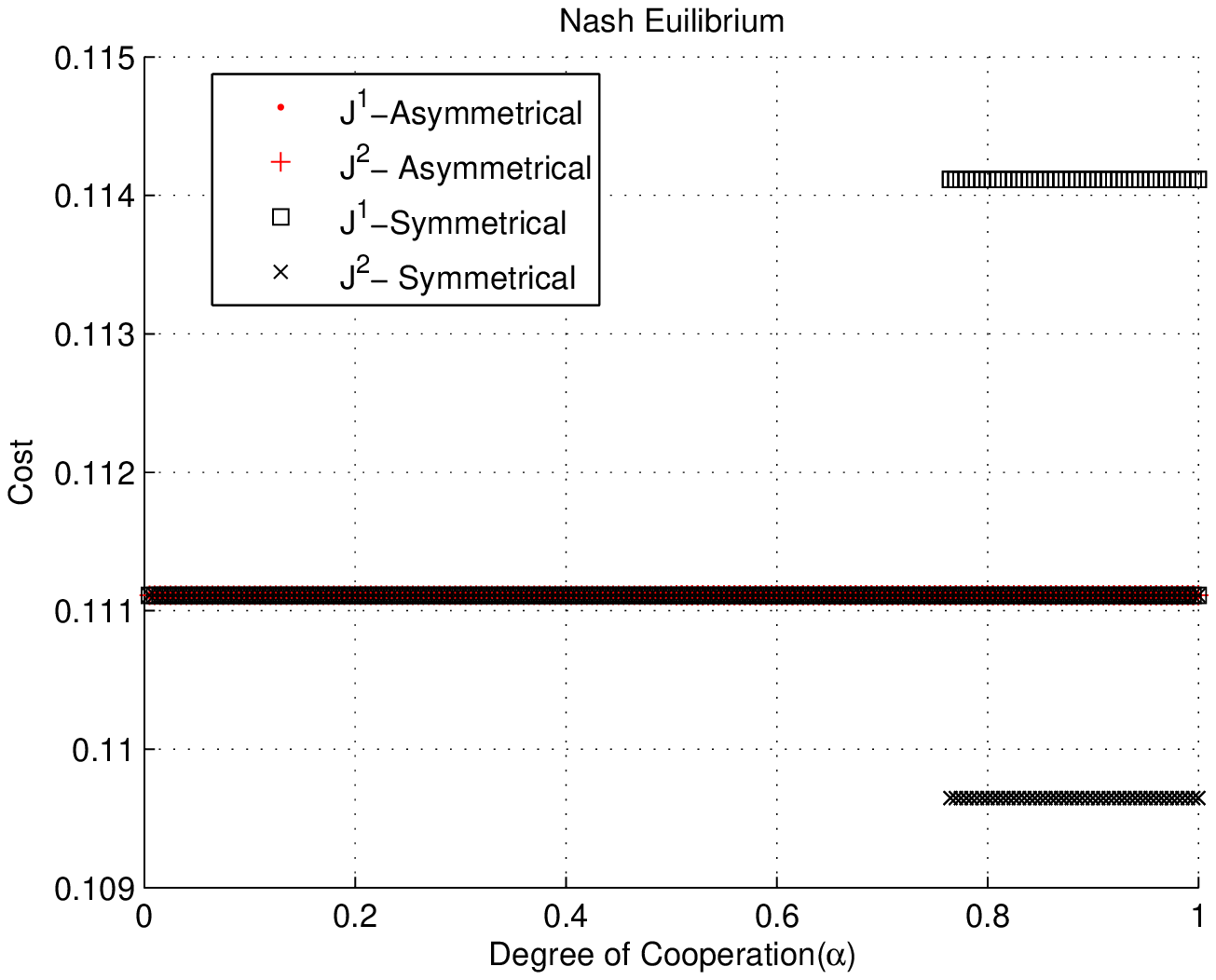}}
\hfill{} \subfigure[Flow values at Nash equilibrium ]{
\label{fig:nashc4:x}
\includegraphics[width=.4\textwidth]{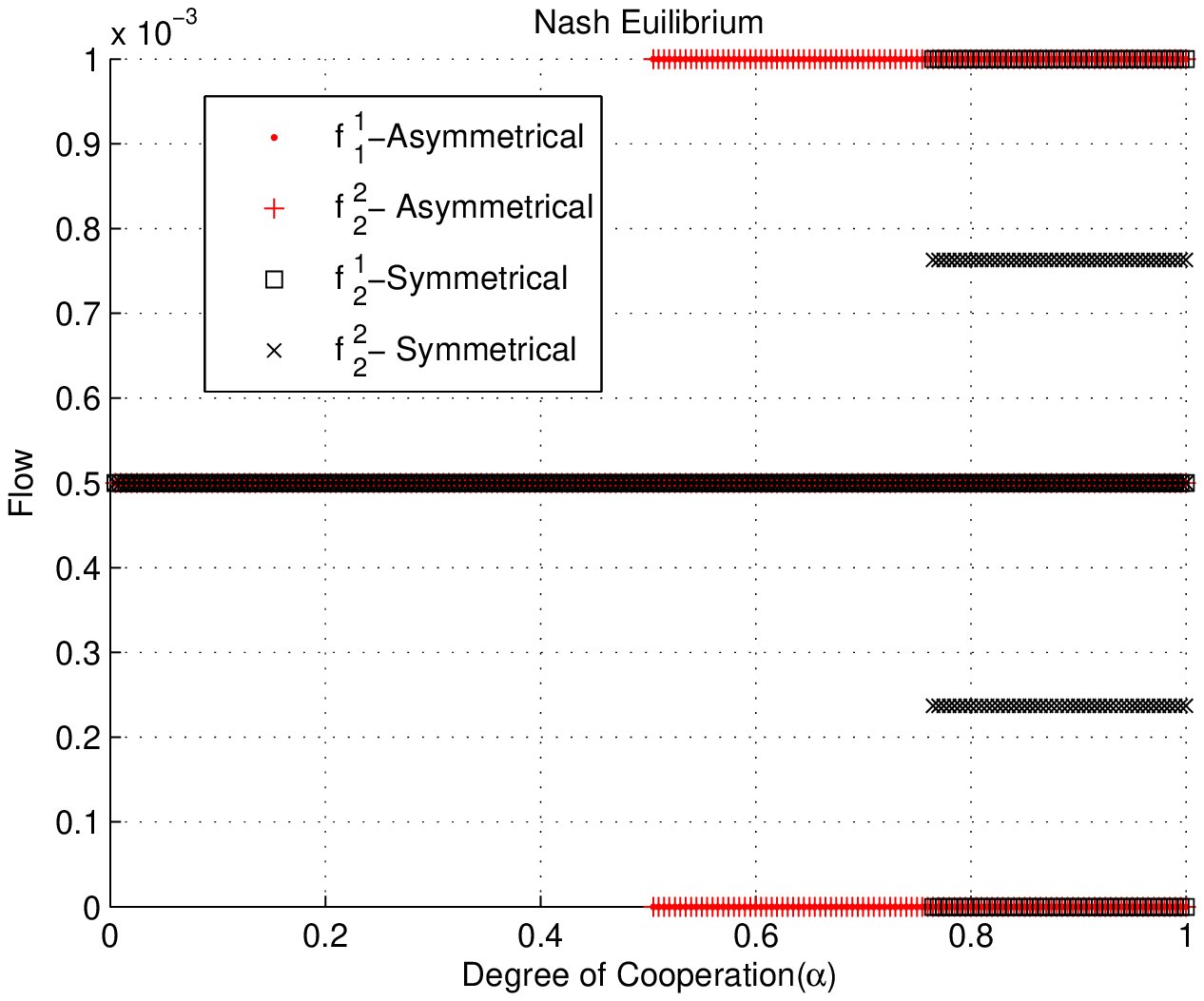}}
{}\caption{Topology : Parallel links, Cost function : M/M/1 Delay,
Parameters : $C_{l_1}=0.001, C_{l_2}=0.001, r^1=1, r^2=1$,
Cooperation: \{ Symmetrical: $\alpha^1=\alpha^2$, Asymmetrical:
$0\leq\alpha^1\leq1, \alpha^2=0 $\}.}\label{fig:nashc4} {}
\end{figure}

{ \emph{Experiment 4) Parallel links with M/M/1 link cost}}:
 In Fig. \ref{fig:nashc4:J},\ref{fig:nashc4:x}, we plot the cost function and
the flow for both users obtained at Nash equilibrium versus
$\alpha$.

\begin{figure}[tb]
\includegraphics[width=.4\textwidth]{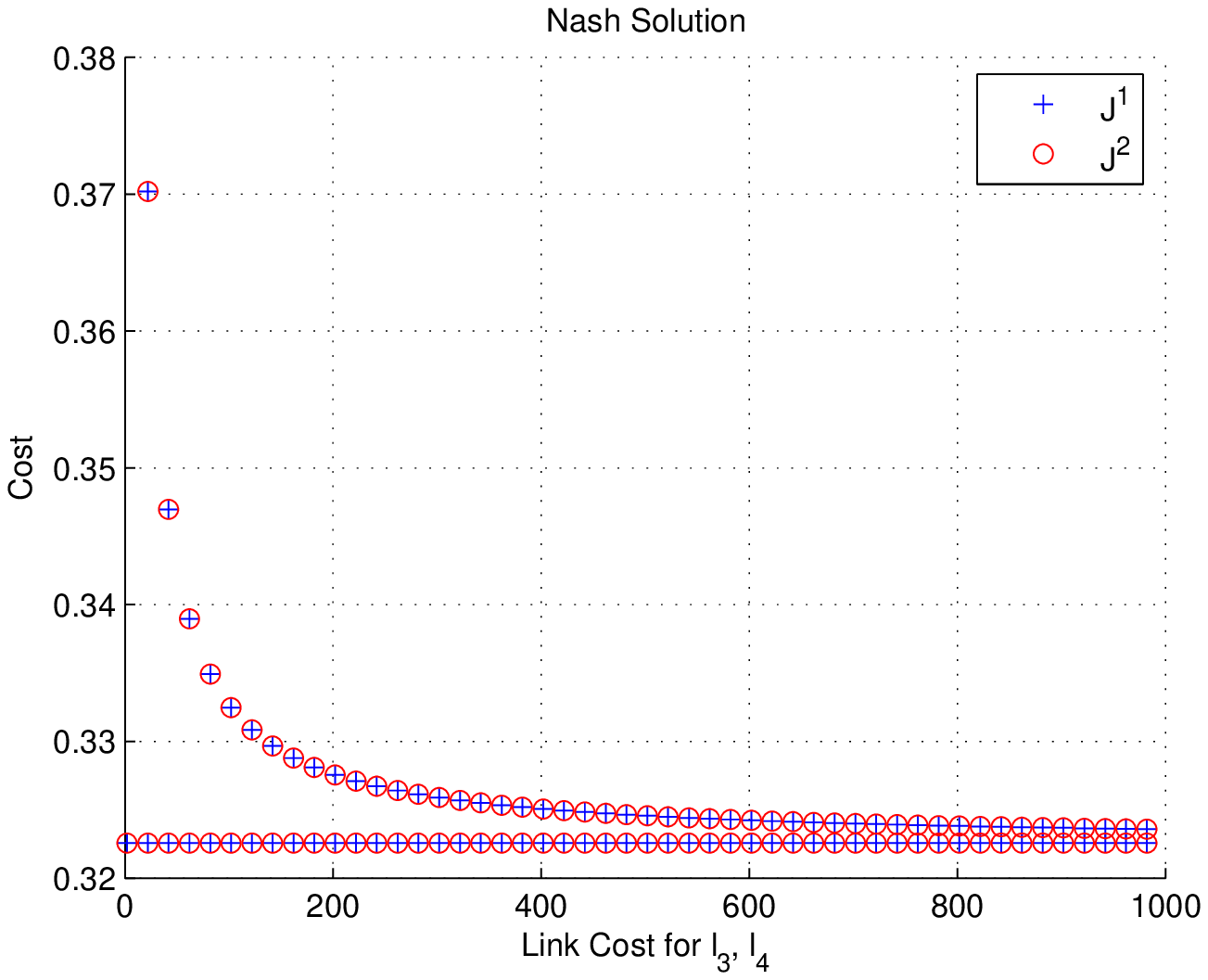}
\caption{Topology : Load balancing, Cost function : Linear,
Parameters : $a_1=4.1,a_2=4.1,d=0.5$,
Cooperation:$\alpha^1=\alpha^2=0.93$ .}\label{fig:braess_load}
\end{figure}

 { \emph{Experiment 5) Load balancing network with linear link cost}}:
We vary the link cost for $l_3$ and $l_4$ by varying the parameter
$c$. More precisely, we increase $c$ from $0$ to $1000$ in the steps
of $20$ and compute Nash equilibrium at each point. In
Fig.\ref{fig:braess_load}, we plot the cost of each user with the
increasing link cost of the link $l_3$ and $l_4$. Note that when the
link cost is high  signifies that link doesn't exit.
We analyze the results obtained from the experimentation done above.
We will be using $\alpha$ and $\alpha^1$ alternatively here for
simplicity as we have fixed $\alpha^2=1$ for asymmetrical case and
$\alpha^1=\alpha^2$ for symmetrical case. The important behavior can
be summarized under following two headings.
\subsection{Non uniqueness of Nash equilibrium}
In Fig. \ref{fig:nashc0}, we observe that there exist
\textbf{multiple Nash equilibria} for both symmetrical case and
asymmetrical case. Note that multiple Nash equilibria is constrained
to some range of cooperation($\alpha$). However there also exist
some range of cooperation where unique solution exist.  We observe
that there exist three Nash equilibrium for some range of
cooperation, two Nash equilibrium at one point and, unique Nash
equilibrium for some range of cooperation. In Fig. [4-6](a,b), we
obtain multiple Nash equilibria as above for some range of
cooperation. In Fig. \ref{fig:nashc2}-\ref{fig:nashc4} although
$\alpha^1=\alpha^2$, due to other parameter being non-symmetrical,
we do not observe a symmetrical plot for "$J^1,J^2$-Symmetrical".
Uniqueness of Nash equilibrium is shown in \cite{orda}, for a
similar situation as in Fig. \ref{fig:nashc2} for selfish user, but
we observe loss of uniqueness when users have some cooperation.

\subsection{ Braess like paradox }

\label{subsec:bras_exa} We also observe a Braess kind of paradox
which is related to performance when additional resource is added to
the system. To understand this, consider the topology of experiment
$1$, i.e., the load balancing network topology. Consider a
configuration where initially link $l_3$ and $l_4$ has very high
cost (effectively doesn't exist) and latter the link cost is reduced
to a low value e.g. $c=0$ and $d=0.5$. This can be interpreted as an
additional resources added to the system. Observe than for the
initial configuration the cost of user $1$ is $J^1=1$ and cost of
user $2$ is $J^2=1$ in experiment 1. However in the latter
configuration which is depicted in Fig. \ref{fig:nashc0:J}, we
observe the cost of user $1$ and $2$ is greater that $1$ at Nash
equilibria. This explains degradation of performance when resources
are increased. A very clearer observation can be made in
Fig.\ref{fig:braess_load} where the link cost for link $l_3$ and
$l_4$ is parameterized. Due to multiple Nash equilibria we see two
curves. The lower curve corresponds to Nash solutions where flow for
each user choose direct link to destination while the upper curve
correspond to mixed strategy solution where a fraction of flow for
each user choose direct link path. Notice that user cost is
improving as the link cost is increasing for the upper curve. Such
paradox is widely studied as \textbf{Braess paradox} in many
literature. Above we identified the existence of Braess paradox in
load balancing network. Now we identify the Braess paradox in
parallel links topology. Consider the parameters as follows,
$C_{l_1}=4.1\, C_{l_2}=4.1 \, r^1=2 \,r^2=1$. Consider the scenario
when initially the link $l_3,l_4$ does not exist, while latter they
are added in the system. In other words, the initially the capacity
$C_3=0, C_4=0$, and latter it is $C_3=10, C_4=10$. Note that when
$C_3=0, C_4=0$, flow at Nash equilibrium is trivially
$f_{l_1}=1,f_{l_2}=1$. In the following, we consider two scenarios
of degree of cooperation :

{\it Only one user is Cooperative :} The degree of Cooperation taken
in this case is $\alpha^1=0.93, \alpha^2=0$. On increasing the
capacity $C_3,C_4$ from $0\rightarrow 10$, the cost functions at
Nash equilibrium are obtained as $J^1=0.952\rightarrow 2.06 $,
$J^2=0.3225\rightarrow 0.909 $ and the flows are $f_{l_1}=2
\rightarrow 0$, $f_{l_2}= 1 \rightarrow 0.0951 $. We also obtain
another Nash equilibrium where the cost functions and the flow
doesn't change from initial state. Note that increasing the capacity
in the network degrades the performance at the first Nash
equilibrium.

{\it Both users are Cooperative :} We repeat the above experiment
with the degree of Cooperation $\alpha^1=0.9, \alpha^2=0.9$. The
cost functions at Nash equilibrium are obtained as $J^1=0.952
\rightarrow 1.247 $, $J^2=0.3225\rightarrow 0.430 $. We again obtain
another Nash equilibrium where the cost functions and the flow
doesn't change from initial state. Note that again increasing the
capacity in the network degrades the performance at the first Nash
equilibrium.
\subsection{Paradox in cooperation} 
In Fig. \ref{fig:nashc0:J}, we observe that $J^1$ has higher cost
than $J^2$. This is intuitive because user $2$ is selfish user while
user $1$ has a varying degree of Cooperation. In particular remark
that  $\alpha^1 \downarrow 0$, $J^1 \uparrow J^2$. But this is not
true for the whole range of Cooperation. Observe in Fig. (3.a) a non
intuitive behavior for some small range of $\alpha^1$ (approximately
$\alpha^1\in(0.87,1)$. Notice that when the degree of cooperation
$\alpha^1$ increases (i.e. increase in its altruism) while other
user be pure selfish $(\alpha^2=0)$, leads to improved cost of user
$1$. This is a paradoxical behavior, we call it \textbf{paradox in
cooperation}. This paradox also exist in case of symmetrical
cooperation (see {\it $J^1$-Symmetrical, $J^2$-Symmetrical }) in the
range of $\alpha$ approximately ($ 0,0.4$). Notice that such paradox
is still observed in Fig. \ref{fig:nashc2}-\ref{fig:nashc6}. Remark
that such paradox exist only when there are multiple equilibria.
\section{Mixed Equilibrium}
\label{sec:mixed_eq}
 The concept of mixed-equilibrium (M.E.) has
been introduced by Harker \cite{Harker} (and further applied in
\cite{dynamic_mixed} to a dynamic equilibrium and in \cite{mixed_eq}
to a specific load balancing problem). Harker has established the
existence of the M.E., characterized it through variational
inequalities, and gave conditions for its uniqueness. We discuss
here the behavior of mixed equilibrium in presence of partial
cooperation. Consider the network $({\cal V, L} )$ shared  by two
types of users: (i) \emph{group users} (denoted by $\cal{N}$) :
these  users have to route a large amount of jobs; (ii)
\emph{individual users}; these users have a single job to route
through the network form a given source to a given destination.
There are infinitely many individual users. For simplicity, we
assume that all individual users have a common source $s$  and
common destination $d$. Let ${\cal P}$ be the set of possible paths
which go from $s$
to $d$. \\
{\bf Cost function}\\
- $J^i: \textbf{F}\rightarrow [0,\infty) $ is the cost function for
each user $i\in \cal{N}$\\
- ${\cal F}_p : {\textbf{F}}\rightarrow [0,\infty) $, is the cost
function of path $p$ for each individual user.\\
The aim of each user is to minimize its cost, i.e., for $i\in
\cal{N}$, $\min_{f^i} J^i({\bf f})$ and for individual user,
$\min_{p\in {\cal P}} {\cal F}^i_p({\bf f})$. Let $f{p}$ be the
amount of individual users that choose path $p$. {}
\begin{definition}
\label{def:mixed} $\textbf{f}\in {\textbf{F}}$ is a Mixed
Equilibrium (M.E.) if \bears \label{eq:mixed_nash}&&\forall i\in
{\cal N}, \forall g^i \textrm{s.t.}(\textbf{f}^{-i},g^i)\in
{\textbf{F}}, \hat{J}^i(\textbf{f})\leq
\hat{J}^i(\textbf{f}^{-i},g^i) \,\\
\label{eq:mixed_war}&&\forall p\in {\cal P},
F_{(p)}(\textbf{f})-A\geq 0; \, (F_{(p)}(\textbf{f})-A)f^i_{(p)}=0
\eears where $A=\min_{p\in {\cal P}} {\cal F}_p({\bf f})$
\end{definition}

\subsection{Mixed equilibrium in parallel links}
In the following proposition, we provide  some closed form of Mixed
equilibrium in parallel links.


\begin{proposition} \label{prop:1}
Consider parallel links network topology (Fig. \ref{fig:para_link})
and M/M/1 delay link cost function. Consider that a Group type user
and Individual type users are operating in this network. The mixed
equilibrium strategy $(f_{l_1}^{1^*},f_{l_2}^{2^*})$ can be given
exactly as follows,\\
\begin{enumerate}
  \item {\it When Both link is used at Wardrop equilibrium:}\\
 $\left\{
  \begin{array}{ll}
(M_1,N_1) &\textrm{if}\,\, a_1<M_1<b_1; \\
otherwise&\\
(0,-cc) &\textrm{if}\,\, r_1<{ \scriptstyle \min\left(r_2+C_2-C_1,
\frac{\alpha(C_2-C_1)+2\alpha r_2}{2\alpha-1}\right)},\\
(r_1,r_1-cc) &\textrm{if}\,\, r_1<{ \scriptstyle
\min\left(\frac{\alpha(C_2-C_1)}
{1-2\alpha},r_2-(C_2-C_1)\right)} , \\
  \end{array}
\right.$ \\where{} \bears M_1={
\scriptstyle\frac{-\alpha(C_2-C_1)+r_1(2\alpha-1)}{2(2\alpha-1)}},\,\,
N_1= { \scriptstyle \frac{(C_1-C_2)(1-\alpha)+(2\alpha-1)r_2}{2(2\alpha-1)}},&&\\
a_1={ \scriptstyle \max(- \frac{C_2-C_1}{2}-\frac{r_2-r_1}{2} ,0
)},\,\, b_1={ \scriptstyle
\min(-\frac{C_2-C_1}{2}+\frac{r_1+r_2}{2},r_1) },&&\\
cc= { \scriptstyle -\frac{C_2-C_1}{2}-\frac{r_2-r_1}{2}},\,\, dd= {
\scriptstyle -\frac{C_2-C_1}{2}+\frac{r_2+r_1}{2}},\,\,\,\,\,&&
\eears

\item {\it When only one link ($1$) is used at  Wardrop equilibrium:}\\

 $\left\{
  \begin{array}{ll}
 (M_2,0) &\,\,\textrm{if}\,\,\, c_1<M_2<r_1; \\
 otherwise&\\
(c_1,0) &\,\,\textrm{if}\,\,\,h(r_1)>0,\\
(r_1,0) &\,\,\textrm{if}\,\,\,h(r_1)<0,\\
 \end{array}
\right.$ \\where $c_1=\max( -\frac{C_2-C_1}{2}-\frac{r_2-r_1}{2} ,0
)$ and $M_2$ is the unique (if there exists) root of the quadratic
equation{} \bears h(x)=ax^2+bx+c=0{} \eears in $[c_1,r_1]$. The
coefficients of the quadratic equation are $ {\scriptstyle
a=((C_1-C_2+r_2)(1-\alpha)-\alpha r_2);
\,\,b=(C_1(1-\alpha)(2(C_2-r_2-r_1)}\\ {\scriptstyle
+2(C_2-r_2))+2\alpha r_2C_1);\,\, c= C_1(1-\alpha)[
(C_2-r_1-r_2)^2}\\{\scriptstyle -C_1(C_2-r_2)]-\alpha r_2C_1^2}.$

\item {\it When only one link ($2$) is used by Wardrop user:}

$ \left\{
  \begin{array}{ll}
(M_3,r_2) &\,\,\textrm{if}\,\,\, 0<M_3<d_1; \\
otherwise&\\
(0,r_2) &\,\,\textrm{if}\,\,\,h(0)>0,\\
(d_1,r_2) &\,\,\textrm{if}\,\,\,h(0)<0,\\
 \end{array}
\right.$ \\where $d_1=\min( -\frac{C_2-C_1}{2}+\frac{r_2+r_1}{2}
,r_1 )$ and $M_3$ is the unique root(if there exist) of the
quadratic equation {}\bears g(x)=ax^2+bx+c=0 {}\eears in $[0,d_1]$.
The coefficients of the quadratic equation are $ {\scriptstyle
a=((C_1-C_2+r_2)(1-\alpha)-\alpha r_2);\,\,
b=(C_1(1-\alpha)}\\{\scriptstyle
(2(C_2-r_2-r_1)+\\2(C_2-r_2))+2\alpha r_2C_1);\,\, c=
C_1(1-\alpha)}\\{\scriptstyle [ (C_2-r_1-r_2)^2-C_1(C_2-r_2)]-\alpha
r_2C_1^2 }.${}
\end{enumerate}
\end{proposition}
{}
\begin{Prf}
We first state the general condition for the mixed equilibrium to
exist. Based on link uses, there are $3$ scenarios when Wardrop
conditions can be met for equilibrium to exist. We individually
state each of them and then we establish the conditions for
equilibria.

For link cost to be finite the link flow must satisfy the flow
constraint $f_{l_1}<C_1,\,\, f_{l_2}<C_2$. From this we obtain the
general condition $r_1+r_2< C_1+C_2$. Equilibria can be attained in
the following conditions:
\begin{enumerate}
  \item {\it When both link is used by Wardrop users:}
Wardrop users utilize both the links, i.e., $f^2_{l_1}>0, \,
f^2_{l_2}>0$, implies cost of both links are same, i.e.,
$T_{l_1}(f_{l_1})=T_{l_2}(f_{l_2})$ (we use $T_{l_1}(f_{l_1})$
instead of ${\cal F}_{l_1}(f_{l_1})$  from def. \ref{def:mixed}).
From
 $T_{l_1}(f_{l_1})=T_{l_2}(f_{l_2})\Rightarrow
f_{l_2}^{2}= -cc+f_{l_1}^{1},$ $0< f_{l_1}^{1}<r_1,$ and
$0<f_{l_2}^{2}<r_2$ imply that $a_1\leq f_{l_1}^{1^*} \leq b_1$,
where $a_1= \max(cc,0)$, $b_1=\min(dd,r_1)$, $cc=-\frac{C_2-C_1}
{2}-\frac{r_2-r_1}{2}$ and $dd=
-\frac{C_2-C_1}{2}+\frac{r_2+r_1}{2}$. Thus the necessary conditions
for equilibrium to exist reduces to $r_1+r_2> |C_1-C_2|$ by noting
$cc<r_1$ and $dd>0$.
Thus the equilibrium strategy $(f_{l_1}^{1^*},f_{l_2}^{2^*})$ is given by\\
 $\left\{
  \begin{array}{ll}
(M_1,N_1) &\textrm{if}\,\, a_1<M_1<b_1; otherwise,\\
(0,-cc) &\textrm{if}\,\, r_1<{\scriptstyle \min\left(r_2+C_2-C_1,
\frac{\alpha(C_2-C_1)+2\alpha r_2}{2\alpha-1}\right)},\\
(r_1,r_1-cc) &\textrm{if}\,\, r_1<{ \scriptstyle
\min\left(\frac{\alpha(C_2-C_1)}
{1-2\alpha},r_2-(C_2-C_1)\right)} , \\
  \end{array}
\right.$ where \bears M_1={
\scriptstyle\frac{-\alpha(C_2-C_1)+r_1(2\alpha-1)}{2(2\alpha-1)}},\,\,
N_1= { \scriptstyle
\frac{(C_1-C_2)(1-\alpha)+(2\alpha-1)r_2}{2(2\alpha-1)}}.\eears

Note that $J^1(f_{l_1}^{1},f_{l_2}^{2})$ is strict convex in the
range $0 < f_{l_1}^{1}< r_1,0 < f_{l_2}^{2^*} < r_2$( by definition
of M/M/1 cost function). It can be directly inferred that if the
equilibrium point $(M_1,N_1)$ satisfies the condition $a_1<M_1<b_1$,
(it is an interior point) there exist atmost one equilibrium.

Otherwise when there is no interior equilibrium point, there may
exist equilibrium at $fl_1^1=0$ or $fl_1^1=r_1$, i.e at point
$(0,-cc)$ or at point ($r_1,r_1-cc$) (since
$Tl_1(f_{l_1})=Tl_2(f_{l_2})$ implies $f_{l_2}^2=-cc+fl_1^1$). The
point $(0,-cc)$ can be an equilibrium point only when
$a_1=\max(0,cc)=0$ and $J'^1(0,cc)>0$. This directly implies $r_1<
r_2+(C_2-C_1)$, and  $r_1< \frac{\alpha(C_2-C_1)+2\alpha
r_2}{2\alpha-1}$ respectively. Combining these, we get $r_1 < \min
\left\lbrace r_2 +(C_2-C_1),\frac{\alpha(C_2-C_1)+2\alpha
r_2}{2\alpha-1} \right \rbrace$. Following the similar steps we can
directly obtain that point $(r_1,r_1-cc)$ can be an equilibrium
point when $r_1< \min \left\lbrace \frac{\alpha (C_2-C_1)}{1-2
\alpha}, r_2- (C_2-C_1) \right\rbrace
$.\\

\item {\it When only one link (link $1$) is used by Wardrop user:}\\
In this case, Wardrop users utilize only link $1$, i.e.,
$f_{l_2}^2=0$. This directly implies $ T_{l_1}(f_{l_1})\leq
T_{l_2}(f_{l_2}) \Rightarrow f_{l_1}^1 \leq cc$ (from wardrop
condition). Combining the above with positive flow condition $0\leq
f_{l_1}^{1^*} \leq r_1$, we obtain  $0\leq f_{l_1}^{1^*} \leq c_1$,
where $c_1=\min\left\lbrace cc,r_1 \right \rbrace$. Since $c_1$ must
be greater than $0$, the necessary condition for equilibrium to
exist reduces to $r_1-r_2 \geq C_1-C_2$.
Further the equilibrium strategy $(f_{l_1}^{1^*},f_{l_2}^{2^*})$ is given by\\
 $ \left\{
  \begin{array}{ll}
(M_2,0) &\,\,\textrm{if}\,\,\, 0<M_2<c_1; \\
otherwise,&\\
(0,0) &\,\,\textrm{if}\,\,\,h(0)>0,\\
(c_1,0) &\,\,\textrm{if}\,\,\,h(0)<0,\\
 \end{array}
\right.$\\ where $M_2$ is the unique root of quadratic equation
$h(x)=ax^2+bx+c$.
 Let $x_1=\frac{-b+\sqrt{D}}{2a},x_2=\frac{-b-\sqrt{D}}{2a}$ are the roots of the Quadratic equation
 $h(x)=0$, where  $ a=(C_1-C_2-r_2)(1-\alpha)+\alpha r_2;\,\,
b=2(1-\alpha)[(C_1-r_2)(2(C_2-r_2)+r_1)]+2\alpha r_2 (C_2-r_1);\,\,
c= (1-\alpha)(C_1-r_2)[(C_2-r_1)^2
-(C_2-r_1)(C_1-r_2)-r_1(C_1-r_2)]+\alpha r_2 (C_2-r_1)^2;\,\,
D=b^2-4ac.$\\
The quadratic equation $h(x)=0$ will have unique solution in the
range $0<f_{l_1}^1<r_1$ because $J'^1(f_{1}^1,0)$ is strict convex
in the range $0<f_{l_1}^1<r_1$ ( by definition of M/M/1 cost
function). Hence there can be atmost one equilibrium point
satisfying $0<M_2<c_1$(i.e single interior point).

Otherwise when there is no interior equilibrium point, there may
exist equilibrium at $fl_1^1=0$ or $fl_1^1=r_1$, i.e., at point
$(0,0)$ or at point ($r_1,0$). The point $(0,0)$ can be an
equilibrium point only when $J'^1(0,0)>0$, i.e., $h(0)>0$. Similarly
point ($c_1,0$) can be equilibrium point only when $J'^1(0,0)<0$,
i.e., $h(0)<0$.

\item {\it When only one link ($2$) is used by Wardrop user:}\\
In this case Wardrop users utilize only link $2$, i.e.,
$f_{l_2}^2=r_2$. Following the similar steps as before,
 we obtain
$d_1\leq f_{l_1}^{1^*} \leq r_1$, where $d_1=\max\left\lbrace dd,0
\right \rbrace$. Since $d_1$ must be less than $r_1$, the necessary
condition for equilibrium to exist reduces to $r_1-r_2 \leq
C_2-C_1$.

Further the equilibrium strategy $(f_{l_1}^{1^*},f_{l_2}^{2^*})$ is given by\\

$ \left\{
  \begin{array}{ll}
(M_3,r_2) &\,\,\textrm{if}\,\,\, d_1<M_3<r_1; \\
otherwise&\\
(0,r_2) &\,\,\textrm{if}\,\,\,h(r_2)>0,\\
(d_1,r_2) &\,\,\textrm{if}\,\,\,h(r_2)<0,\\
 \end{array}
\right.$ \\where  $M_3$ is the unique root(if there exist) of the
quadratic equation $g(x)=ax^2+bx+c$ in $d_1< f_{l_1}^1 < r_1$. Let
$x_1=\frac{-b+\sqrt{D}}{2a},x_2=\frac{-b-\sqrt{D}}{2a}$ are the
roots of the Quadratic equation $ g(x)=0$, where $
a=((C_1-C_2+r_2)(1-\alpha)-\alpha r_2); \,\,b=(1-\alpha)[
4C_1(C_2-r_1-r_2)+2r_1C_1]- 2\alpha r_2C_1);\,\, c=
(1-\alpha)[(C_2-r_1-r_2+C_1)C_1 (C_2-r_2-r_1)- r_1C_1^2]+\alpha
r_2C_1^2; \,\,
 D=b^2-4ac.$
\\
The quadratic equation $g(x)=0$ will have unique solution in the
range $0<f_{l_1}^1<r_1$ because $J'^1(f_{1}^1,r_2)$ is strict convex
in the range $0<f_{l_1}^1<r_1$ ( by definition of M/M/1 cost
function). Hence there can be atmost one equilibrium point
satisfying $d_1<M_3<r_1$(i.e single interior point).

Otherwise when there is no interior equilibrium point, there may
exist equilibrium at $fl_1^1=0$ or $fl_1^1=r_1$, i.e., at point
$(0,r_2)$ or at point ($r_1,r_2$). The point $(0,r_2)$ can be an
equilibrium point only when $J'^1(0,r_2)>0$, i.e., $g(r_2)>0$.
Similarly point ($r_1,r_2$) can be equilibrium point only when
$J'^1(0,r_2)<0$, i.e., $g(r_2)<0$.

\end{enumerate}
\end{Prf}

\begin{corollary}
Consider the symmetric parallel links, i.e., $(C_1=C_2=C,
r_1=r_2=r)$ network with M/M/1 delay link cost function. In a mixed
user setting the mixed equilibrium strategy
($(f_{l_1}^{1^*},f_{l_2}^{2^*})$) can be given by $\left\{
  \begin{array}{ll}
(\frac{r}{2},\frac{r}{2}) &\,\,\textrm{when}\,\, r_1>f_{l_1}^1>0,r_2>f_{l_2}^2>0 \\
(0,0) &\,\,\textrm{when}\,\,0\leq f_{l_1}^1 \leq r_1,f_{l_2}^2=0,\,\, \textrm{if}\,\, \alpha \geq 0.5\\
(r,r) &\,\,\textrm{when}\,\,0\leq f_{l_1}^1 \leq r_1,f_{l_1}^2=0,\,\, \textrm{if}\,\, \alpha \geq 0.5\\
 \end{array}
\right.$
\end{corollary}
{}
\begin{Prf}

Consider the symmetric case when $C_1=C_2=C$, $r_1=r_2=r$. The
general condition thus reduces to $r<C$ from prop. \ref{prop:1}.
Equilibrium can be attained under the following scenario based on
link uses.

\begin{enumerate}
  \item {\it When both link is used by Wardrop users:}\\
Wardrop users utilizes both the links, i.e., $f_{l_1}^2>0,
f_{l_2}^2>0$, implies cost function of both the links are same,
i.e., $T_{l_1}(f_{l_1})=T_{l_2}(f_{l_2})$. From
$T_{l_1}(f_{l_1})=T_{l_2}(f_{l_2})\Rightarrow f_{l_2}^{2}=
f_{l_1}^{1},$ $0< f_{l_1}^{1}<r,$ and $0<f_{l_2}^{2}<r$, implies
that necessary condition for equilibrium to exist are always
satisfied. Further the equilibrium strategy
$(f_{l_1}^{1^*},f_{l_2}^{2^*})$
 is given by $(\frac{r}{2},\frac{r}{2})$ which can be directly
 obtained from prop. (\ref{prop:1}.1).
 \\
  \item {\it When only one link (link $1$) is used by Wardrop user:}
In this case, Wardrop users utilize only link $1$, i.e.,
$f_{l_2}^2=0$. This directly implies $ T_{l_1}(f_{l_1})\leq
T_{l_2}(f_{l_2}) \Rightarrow f_{l_1}^1 \leq 0$ (from wardrop
condition). Combining the above with positive flow condition $0\leq
f_{l_1}^{1^*} \leq r_1$, we obtain  $ f_{l_1}^{1^*} = 0$. This
suggests that equilibrium point can be given by
$(f_{l_1}^{1^*},f_{l_2}^{2^*})=(0,0)$ if there exist.

Note that $(0,0)$ is the boundary point solution. If $J'^1(0,0)\geq
0$ (Nash solution of user $1$) then the equilibrium point is given
by $(0,0)$. ${J^1}'(f_{l_1}^{1},0)$ can be expressed
 as $\frac{P(x)}{Q(x)}$, where \bears &&P(x)= ax^2+bx+c \eears
$a=r(2\alpha-1);\,\, b=2(C-r)(2(C-r)(1-\alpha)+r);\,\,
c=(2\alpha-1)r(C-r)^2;\,\, D=16 (C-r)^2 (1-\alpha)C
[(C-r)(1-\alpha)+\alpha r]$ and $Q(x)>0$ for all $x$,  thence
$J'^1(0,0) \geq 0 \Rightarrow c \geq 0 \Leftrightarrow \alpha \geq
0.5$.

  \item {\it When only one link ($2$) is used by Wardrop user:}
In this case, Wardrop users utilize only link $2$, i.e.,
$f_{l_2}^2=r$. This directly implies $ T_{l_1}(f_{l_1})\geq
T_{l_2}(f_{l_2}) \Rightarrow f_{l_1}^1 \geq r$ (from Wardrop
condition). Combining the above with positive flow condition $0\leq
f_{l_1}^{1^*} \leq r_1$, we obtain  $ f_{l_1}^{1^*} = r$. This
suggests that equilibrium point can be given by
$(f_{l_1}^{1^*},f_{l_2}^{2^*})=(r,r)$ if there exist.

Remark that this case is symmetrical to case when only link $1$ is
used. Hence we can directly infer the condition for equilibrium
point to exist. The equilibrium point point $(r,r)$ exist, when
$\alpha \geq 0.5 $.

\end{enumerate}

\end{Prf}

%

\begin{figure}[tb] {}
\begin{center}
{\includegraphics[width=60mm]{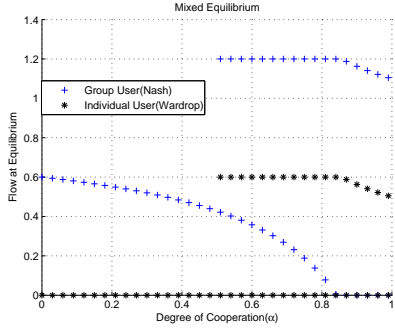}}
{}\caption{Topology : Parallel links, Cost function : M/M/1 Delay,
Parameters : $C_{l_1}=4, C_{l_2}=3, r^1=1.2, r^2=1$ .}
\label{fig:mixed_f_LCPN}
\end{center}
\end{figure}

In Fig. (\ref{fig:mixed_f_LCPN}), we depict the mixed equilibrium
strategy(flow) for the varying degree of cooperation$(\alpha)$.
Observe the loss of uniqueness of mixed equilibrium in presence of
partial cooperation. It is known to have unique equilibrium in the
network setting with finitely many selfish users\cite{mixed_eq}.
Remark that we have already shown in the previous section that there
exist multiple Nash equilibria in presence of partial cooperation.
Due to space limitation we illustrate this behavior with only
parallel links topology and M/M/1 cost function. However we identify
a similar remark from other configuration also.

\section{Existence and uniqueness of Equilibrium}
\label{sec:unique} Having noted the existence of multiple Nash
equilibrium in sec.\ref{sec:num_ex} using various examples, we here
establish the conditions under which unique nash equilibrium exist.
Uniqueness of Nash equilibrium is shown in \cite{orda} in case of
non-cooperative games for parallel links topology. Under some
condition, uniqueness is shown for general topology also. In this
section we follow the similar structure to establish the uniqueness
for parallel links topology in case of our setting of user
cooperation.

We follow some assumptions on the cost function $J^i$ same as in
\cite{orda}.
 {}
\begin{assumption}:\\ \label{ass:G}
\begin{tabular}{cl}
G1: & ${J}^i({\bf{f}})=\sum_{l\in {\cal L}} \hat{J}^i_l(f_l))$. Each $\hat{J}^i_l$ satisfies:\\
G2: & ${J}^i_l$:$[0, \infty) \rightarrow (0,\infty]$ is continuous function.\\
G3: & ${J}^i_l$: is convex in $f^j_l$ \,\, for $j=1,...|{\cal I}|$. \\
G4: & Wherever finite, ${J}^i_l$ is continuously differentiable\\
& in $f^i_l$, denote $ K^i_l=\frac{\delta \hat{J}^i_l}{\delta f^i_l}$.\\

\end{tabular}\\ {}
\end{assumption}
Note the inclusion of $+\infty$ in the range of $\hat {J}^i_l$,
which is useful to incorporate implicitly and compactly and
additional constraints such as link capacities. Also note that the
assumption $G3$ is stronger than in \cite{orda}.

Function that comply with these general assumptions, we call type
$G$ function. 
For selfish user operating on parallel links NEP is shown to exist
in \cite{orda} with the function which comply with the type $G$
function.

We shall mainly consider cost functions that comply with the
following assumptions:
%
{}
\begin{assumption}:\\\label{ass:B}
\begin{tabular}{cl}
B1: & ${J}^i({\bf{f}})=\sum_{l\in {\cal L}} f^i_l T_l(f_l))$\\
B2: & $T_l: [0, \infty) \rightarrow (0,\infty]$.\\
B3: & $T_l(f_l)$ is positive, strictly increasing and convex.\\
B4: & $T_l(f_l)$ is continuously differentiable.\\
\end{tabular}\\ {}
\end{assumption}

Functions that comply with these assumptions are referred to as {\it
type}-{\bf B} functions.
 {}
\begin{remark} In Assumption {\bf B1},
$T_{l}(f_{l})$ is the cost per unit of flow (for example mean delay)
on the link $l$, for the total utilization, $f_{l} = \sum_{i \in
{\cal I}}f_l^i $, of that link. Note that if $T_l(f_l)$ is the
average delay on link $l$, it depends only on the total flow on that
link. The average delay should be interpreted as a general
congestion cost per unit of flow, which encapsulates the dependence
of the quality of service provided by a finite capacity resource on
the total load $f_l$ offered to it.
\end{remark}

A special kind of type-B cost function is that which corresponds to
an M/M/1 link model. In other words, suppose that\\
\begin{tabular}{cl}\label{ass:c}
C1: & $\hat{J}^i(f^i_l,f_l) = f^i_l\dot T_l(f_l)$ is
a type-B cost function.\\
C2: & $T_l= \left \{
\begin{array}{ll}
\frac{1}{C_l-f_l} & f_l< C_l \\
\infty & f_l> C_l
\end{array}
\right.$. \\
&Where $C_l$ is the capacity of the link $l$.
\end{tabular}\\
Function that comply with these requirements are referred to as
type-\textbf{C} functions. Such delay functions are broadly used in
modeling the behavior of the links in computer communication
networks \cite{Gallager},\cite{Kleinrock}.

\subsection{Parallel links network topology}
\label{sec:parallel_uniqueness} In this section we study the special
case where the users from set ${ \cal I}$ shares a set of parallel
communication links ${\cal L}=\{1,2.... L\}$ interconnecting a
common source node to a common destination node. In \cite{orda},
uniqueness of Nash equilibrium is shown for the selfish users (when
user do not cooperate in managing the communication link) in
parallel links, where the cost functions ($J^i(\mathbf{f})$) of
users are assumed to hold assumption \ref{ass:B}. However this is
not true when the users have cooperation in degree as defined in
sec.(\ref{sec:coop}). We observe that assumption \ref{ass:B} is not
sufficient to guarantee unique Nash equilibrium in our setting. It
is a harder problem to characterize system behavior for general
degree of cooperation. Hence we consider a special case of
cooperation where a user cooperative with similar cooperation with
all other users i.e. \bears
\hat{J}^i(\textbf{f})=(1-\alpha^i)J^i(\textbf{f})+\alpha^i\sum_k
J^k(\textbf{f}) {}\eears Consider the cost function of type
\ref{ass:B}. The cost function of each user on link $l$ is given by
\begin{eqnarray*}
\hat{J}^i_l(\mathbf{f})&=& ((1-\alpha^i) f_l^i +\alpha^if_l^{-i})
T_l(f_l)\\
&=& ( (1-\alpha^i )f_l+(1- 2\alpha^i) f_l^{-i})T_l(f_l)
\end{eqnarray*}

Existence problem in the case of Nash equilibrium for the cost
function $\hat{J}^i_l(\textbf{f})$ can be directly studied as in
\cite{orda}.

Note that in case of $\alpha^i < 0.5$ for all $i\in {\cal I}$, the
uniqueness of Nash equilibrium is guaranteed from E. Orda et
al.\cite{orda}. Note that when $\alpha^i < 0.5$, the function
$K^i_l(f^{-i}_l,f_l)$ is strictly increasing function in $f_l^{-i}$
and $f_l$.

Uniqueness of Nash equilibrium can be also observed in case of
All-positive flow in each link.  By All-positive flow we mean that
each user have strictly positive flow on each link of the network.%

 The following result establishes the uniqueness of Nash Equilibrium in case of
positive flow. {}
\begin{thm}\label{thm:positiveflow} Consider the cost function of type \ref{ass:B}. Let $\bf{\hat
f}$ and $\bf{f}$ be two Nash equilibria such that there exists a set
of links $\overline{{\cal L}}_1$ such that $\{f_l^i>0 \mbox{ and }
\hat f_l^i>0, i\in{\cal I}\}$ for $l\in\overline{{\cal L}}_1$, and
$\{f_l^i=\hat f_l^i=0, i\in{\cal I}\}$ for $l \not\in\overline{{\cal
L}}_1$. Then $\bf{\hat f}=\bf{f}.$
\end{thm}

\begin{Prf}
Let $\textbf{f} \in F$ and $\hat{\textbf{f}}\in F$ be two NEP's. As
observed $\textbf{f}$ and $\hat{\textbf{f}}$ satisfy the Kuhn-Tucker
condition. We rewrite the Kuhn-Tucker condition in terms of
$f^{-i}_l,f_l$ as below,
 \bears
\label{eq:4}K^i_l(f^{-i}_l, f_l)&\geq& \lambda^i
;K^i_l(f^{-i}_l, f_l)= \lambda^i \,\, \textrm{if} \,\,f^i_l>0\,\, \forall i,l\\
\label{eq:5}K^i_l(\hat{f}^{-i}_l, \hat{f}_l)&\geq& \lambda^i
;K^i_l(\hat{f}^{-i}_l, \hat{f}_l)= \lambda^i \,\,\textrm{if}\,\,
\hat{f}^i_l>0 \,\, \forall i,l \eears The above relation and the
fact that $K^i_l(.; .)$ is increasing in both of is argument will be
used below to establish that $\mathbf{f}=\hat{\mathbf{f}}$ i.e.
$f^i_l=\hat{f}^i_l$ for every $l,i$. The first step is to establish
that $f_l=\hat{f}_l$ for each link $l$. To this end, we prove that
for each $l$ and $i$, the following relation holds: \bear
\label{eq:6}\{ \hat{\lambda}^i \leq \lambda^i, \hat{f}_l \geq f_l
\}&& \,\,\textrm{implies that}\,\, \hat{f}^{-i} \leq f^{-i},\\
\label{eq:7}\{ \hat{\lambda}^i \geq \lambda^i, \hat{f}_l \leq f_l
\}&& \,\,\textrm{implies that}\,\, \hat{f}^{-i} \geq f^{-i}. \eear

We shall prove (\ref{eq:6}), since (\ref{eq:7}) is symmetric. Assume
that $\hat{\lambda}^i\leq \lambda$ and $\hat{f}_l \geq f_l$ for some
$l$ and $i$. For $f^{i}_l>0$ together with our assumptions imply
that: \bear \label{eq:8} K^i_l(\hat{f}^{-i}_l,f_l)= \hat{\lambda}^i
\leq {\lambda}^i \leq K^i_l({f}^{-i}_l,f_l)\leq
K^i_l(\hat{f}^{-i}_l,\hat{f}_l), \eear where the last inequality
follows from the monotonicity of $K^i_l$ in its second argument. Now
, since $K^i_l$ is nondecreasing in its first argument, this implies
that $f^{-i}_l \leq f^i_l$, and (\ref{eq:6}) is established.

Let ${\cal L}_1 = \{ l: \hat{f}_l>f_l\}$. Also denote ${\cal I}_a=\{
i: \hat{\lambda}^i >\lambda^i \}$, ${\cal L}_2 = {\cal L} -{\cal
L}_1 = \{ l: \hat{f}^l \leq f_l\}$. Assume that ${\cal L}_1$ is non
empty. Recalling that $\sum_l \hat{f}^{-i}_l= \sum_l
f^{-i}_l=r^{-i}$, it follows from (\ref{eq:7}) that for every $i$ in
${\cal I}_a$, {\small \bears \hspace{-3mm}\sum_{l\in{\cal
L}_1}\hspace{-1mm} \hat{f}^{-i}= r^{-i}-\sum_{l\in{\cal
L}_2}\hspace{-1mm} \hat{f}^{-i} \leq r^{-i}-\sum_{l\in{\cal
L}_2}\hspace{-1mm} {f}^{-i} =\sum_{l\in{\cal L}_1}\hspace{-1mm}
{f}^{-i}, \,\,\, i\in {\cal I}_a. \eears }

From (\ref{eq:6}), we know that , $\hat{f}^{-i}_l \leq f^{-i}_l$ for
$l \in {\cal L}_1$ and $i\notin {\cal I}_a$, it follows that :
\bears \sum_{l\in {\cal L}_1} \hat{f}_l= \sum_{l\in {\cal
L}_1}\frac{\sum_{i\in {\cal I}} \hat{f}^{-i}_l}{{\cal I } -1} \leq
\sum_{l\in {\cal L}_1}\frac{\sum_{i\in {\cal I}} {f}^{-i}_l}{{\cal I
} -1} = \sum_{l\in {\cal L}_1} {f}_l\eears

This inequality obviously contradicts our definition of ${\cal
L}_1$. Which implies that ${\cal L}_1$ is an empty set. By symmetry,
it may also be concluded that the set $\{ l: \hat{f}_l < f_l\}$ is
also empty. Thus, it has been established that: {} \bear
\label{eq:11}\hat{f}_l= f_l \,\,\,\textrm{for every }\,\,\, l\in
{\cal L}. \eear We now show that $\hat{\lambda}^i= \lambda^i$ for
each user $i$. To this end, note that (\ref{eq:4}) may be strengthen
as follows: \bear\nonumber \{ \hat{\lambda}^i < \lambda^i,
\lambda{f}_l=f_l\} \,\,\,
\textrm{implies that either} && \\
\hat{f}^{-i}_l< f^{-i}_l \,\, \textrm{or}\,\,
\label{eq:12}\hat{f}^{-i}_l=f^{-i}_l=0. &&\eear Indeed if
$f^{-i}_l=0$, then the implication is trivial. Otherwise, if
$f^{-i}_l>0$, it follows similar to (\ref{eq:8}) that
$K^i_l(\hat{f}^{-i}_l,\hat{f}_l)$ that $\hat{f}^{-i}_l < f^{-i}_l$
as required. Assume now that $\hat{\lambda}^i<\lambda^i$ for some
$i\in {\cal I}$. Since $\sum_{l\in {\cal L}} \hat{f}^{-i}_l= r^{-i}
>0$, then $f^{-i}_l >0$ for at least one link $l$ and from
(\ref{eq:12}) implies that, $ \sum_{l\in {\cal L}} f^i_l >
\sum_{l\in {\cal L}} \hat{f}^i_l=r^i, $ which contradicts the demand
constraint for user $i$. We, therefore, conclude that
$\hat{\lambda}^i<\lambda^i $ does not hold for any user $i$. A
symmetric argument may be used to show that
$\hat{\lambda}^i=\lambda^i$ for every user $i\in {\cal I}$. Combined
with (\ref{eq:11}), this implies by (\ref{eq:6}) and (\ref{eq:7})
that $\hat{f}^{-i}_l=f^{-i}_l$ for every $l,i$. Again since
$f^{i}_l=f_l-f^{-i}_l$, uniqueness of $f^i_l$ is proved.

\end{Prf}

\subsection{Uniqueness of NEP in general topology}
It is a hard to characterize system behavior for general network
with user's partial cooperation. For selfish users, it is shown that
there exist uniqueness for Nash equilibrium point(NEP) under
Diagonal Strict Convexity in \cite{orda}.

We consider a special case of cooperation where a user cooperates
equally with all other users i.e. {}\bears
\hat{J}^i(\textbf{f})=(1-\alpha^i)J^i(\textbf{f})+\alpha^i\sum_k
J^k(\textbf{f}) \eears Consider the cost function of type
\ref{ass:B}. The cost function of each user on link $l$ can be thus
given by
\begin{eqnarray}{}
\hat{J}^i_l(\mathbf{f})
&=& ((1-\alpha^i) f_l+(1-2\alpha^i) f_l^{-i})T_l(f_l)
\end{eqnarray}

\begin{thm} {}Consider the cost function of type \ref{ass:B}. Let $\bf{\hat
f}$ and $\bf{f}$ be two Nash equilibria such that there exists a set
of links $\overline{{ \cal L}}_1$ such that $\{f_l^i>0 \mbox{ and }
\hat f_l^i, i\in{\cal I}\}$ for $l\in\overline{{\cal L}}_1$, and
$\{f_l^i=\hat f_l^i=0, i\in{\cal I}\}$ for $l\not\in \overline{{\cal
L}}_1$. Then $\bf{\hat f}=\bf{f}.${}
\end{thm}
Under all positive flows assumption, the Kuhn-Tuker conditions for
all $l=(u,v)\in \cal{L}_1$ becomes   {}\bears
&&{\scriptstyle((1-\alpha^i) f_{l}^i+ \alpha^i f_{l}^{-i})
T'_{l}(f_{l})+(1-\alpha^i) T_{l}(f_l)= \lambda^i_{u}-\lambda_v^i }\\
&& {\scriptstyle ((1-\alpha^i) \hat f_{l}^i+ \alpha^i\hat
f_{l}^{-i}) T'_{l}(\hat f_{l})+(1-\alpha^i) T_{l}(\hat f_{l})
=\hat\lambda^i_{u}-\lambda^i_{v} }\eears Summing each of these
equations over $i$, we obtain {} \bears &&{\scriptstyle
H_{uv}(f_{l}):=(\alpha I+1-2\alpha) T'_{l}(f_{l})+I(1-\alpha)
T_{l}(f_l)=\lambda_{u}-\lambda_v}\\
&& {\scriptstyle H_{uv}(\hat f_{l}):=(\alpha I+1-2\alpha)\hat f_{l}
T'_{l}(\hat f_{l})+(1-\alpha) I T_{l}(\hat f_{l})=
\hat\lambda_{u}-\lambda_{v} }\eears Since the function H is strictly
increasing, we follow the same proof of Theorem 3.3 in \cite{orda}
to conclude that $\bf{\hat f}=\bf{f}$.

\section{Concluding Remarks}
\label{sec:conclusion}

This paper is aimed at exploring user performance in routing games
where a finite number of users take into account not only their
performance but also other's user's performance. We have
parameterized the \emph{degree of Cooperation} to capture the user
behavior from altruistic to ego-centric regime. We notice some
strange behaviors. Firstly we show the existence of multiple Nash
equilibria by a simple example of parallel links and load balancing
networks in contrast to the unique Nash equilibrium in case of
selfish users. Moreover, we then explored the mixed user scenario,
which is composed of a finite number of Group type user seeking Nash
equilibrium and infinitely many Individual type users satisfying
Wardrop condition. We illustrate loss of uniqueness of equilibrium
even in mixed users scenario in presence of partial cooperation by
an example for parallel links network. However it is known to have
unique equilibrium in presence of only finitely many selfish users
in similar settings.

Secondly we identify two kinds of paradoxical behavior. We identify
situation where well known Braess paradox occurs in our setting of
cooperation. We show using an example of parallel links network with
M/M/1 link cost that addition of system resources indeed degrades
the performance of all users in presence of some cooperation, while
it is well known that this is not true for  this setting with only
selfish users.

We also identify another type of paradox, paradox in cooperation:
i.e. when a given user increases its degree of cooperation while
other users keep unchanged their degree of cooperation, this may
lead to an improvement in performance of that given user. In extreme
sense a user can benefit itself by adopting altruistic nature
instead of selfishness.

\end{document}